\documentclass[preprint,showpacs,preprintnumbers,amsmath,amssymb,floats]{revtex4-1}
\usepackage{blindtext}
\usepackage{graphicx}
\usepackage[english]{babel}
\usepackage{amsmath}
\usepackage{amssymb}
\usepackage{color}

\newcommand{\be}{\begin{eqnarray}}
\newcommand{\ee}{\end{eqnarray}}
\usepackage[toc,page]{appendix}

\begin{document}
\title{Majoranas in mixed-valence insulators}
\author{C. M. Varma}
\thanks{chandra.varma@ucr.edu \\
\noindent
$^1$ Visiting Professor. \\
Emeritus Distinguished Professor, Physics Department, University of California, Riverside, CA. 92521.}

\address{$^1$Physics Department, University of California, Berkeley, California 94704}

\date{\today}

\begin{abstract}
A physical model for a mixed-valence impurity in a metal must satisfy the Friedel screening theorem 
for both valences. Such a model is shown, following earlier work which showed low energy
singularities in it, to
be supersymmetric, leading to a free Majorana and a phase-shifted Majorana excitation. 
The theory extended approximately to a lattice of mixed-valence 
ions at appropriate filling gives, without fine-tuning the parameters, a protected gapless Majorana fermion band 
across the chemical potential, 
besides the mixed-valence
particle and hole bands separated by gaps. In this situation
the system is  electrically neutral in linear response but has de Haas-van Alphen oscillations.
This is used to explain the recently observed magneto-oscillations 
 in mixed-valence insulators as well as their accompanying low energy thermodynamic 
 and relaxation rate anomalies.
Some predictions to test the validity of the theoretical results are provided, the most 
striking of which is that there should be extensive ground state entropy in such compounds. 
 \end{abstract}
\maketitle

\section{Introduction}
The theory of heavy-fermion and mixed-valence metals and insulators began through understanding   
the Kondo problem using a variational wave-function for the Anderson model \cite{Anderson1961} as a resonance obtained  by constrained hybridization 
of the strongly correlated 
 magnetic orbitals
obeying local quantum-numbers with the nearly free conduction bands \cite{V_Yafet1, V_RMP_MV}. The problem of the lattice,
simply treated as that of conduction electrons hybridizing with a periodic array of such resonances, 
done in a variety of different ways, for example in \cite{V_Yafet1}, \cite{Fulde_HF},  \cite{AuerbachLevin}, \cite{MillisLeeHF}, \cite{V_Weber_Randall}, 
\cite{Rice_Ueda_Gutzwill}, \cite{DMFT_RMP1996},
is also highly successful in understanding experiments in heavy-fermion metals away from any
singularities due the interactions between the correlated orbitals.
The reasons are 
well understood because the excitations are that of a Fermi-liquid obeying Bloch's
counting rules, albeit with quite different
relation \cite{Varma_HF_phen} of Landau parameters to experimental properties from those in He$^3$, the 
canonical Fermi-liquid. In the case of mixed-valence insulators, predicted by 
the Bloch's rules in the 
same theory based on the same model, this happy state of affairs is called into question by some 
recent experiments  \cite{Sebastian2015SmB6, Li2018YbB12, Sebastian2018YbB12, SebastianSmB62020} 
with results which are spectacular and call for a new paradigm. Oscillatory magnetization in a magnetic field $H$ with period proportional 
as in Onsager on metals to $1/H$, and finite value of the
Sommerfeld coefficient $\gamma = C_v/T$ with a possibly singular correction at low temperature,
 are observed at low temperatures in materials which are 
electrical insulators, of the mixed-valence kind. In this paper, I take these results in context of the fact that the essentially 
exact solution \cite{PerakisVRuck, Sire_V_R_G}
of a proper physical model for a mixed-valence impurity has singular low energy excitations, unlike the
local Fermi-liquid \cite{Nozieres_LFL} properties of a Kondo impurity.  I show here that this solution
\cite{PerakisVRuck, Sire_V_R_G}
is equivalent to having  Majorana excitations. The theory approximately extended to a lattice of mixed-valent ions gives an insulator 
obeying the Bloch rules, but one of the branches of Majoranas survives and is in fact
 protected by the insulating gap. The underlying physics of this Majorana is different from the 
 one-dimensional superconductor model of Kitaev \cite{Kitaev2001}. Although the existence of a Majorana is tantamount to 
 breaking U(1) gauge invariance, there is no implication in the present work for superconductivity of any kind. 
 Arguments for magneto-oscillations in superconductors with
 odd frequency even parity pairing \cite{Balatsky_Abrahams}, \cite{Coleman_Miranda} have already been presented by
  Baskaran \cite{Baskaran_Majorana} and by Erten et al. \cite{Erten_Coleman}.
  But the materials under discussion are not superconductors of any kind.
  
  The physics of the singularities in the two-channel Kondo problem 
 \cite{Nozieres_Blandin, AL2, Emery-Kivelson92} and the 
 two impurity Kondo
 problem \cite{Jones_V_Wilkins, AL3_two_imp1992,  SVK} at specially tuned points may also be expressed
in terms of  Majoranas. However, in both these problems, the singularities occur at specially tuned points, so that
the problems though of theoretical interest are not naturally realizable. The mixed-valence
impurity also displays the singularity at a specially tuned point but in a mixed-valence lattice the phenomena occurs over a range
of parameters and fine tuning is not required.
 
 What is the difference between a mixed-valence impurity and a Kondo impurity? 
 For a mixed-valence impurity,
 two different charge states of the impurity are nearly degenerate at the chemical potential 
 while the third is far away in energy \cite{V_RMP_MV}  \cite{Haldane1977_mixedval} \cite{CMVcorrIns}. For a Kondo impurity, the Hartree-Fock resonance for one charge state is
 below the chemical potential while its adjacent charge states on both sides have resonances far above the chemical potential \cite{Anderson1961}. 
 While weak breaking
  of particle-hole symmetry in the Anderson model  \cite{Anderson1961} is known to be irrelevant \cite{PerakisV, PerakisVRuck},
   mixed valence requires strong particle-hole asymmetry. In that case, both charge 
 states must be locally screened, which means by the Friedel theorem on local charge neutrality that there should be 
 a phase shift of $\pi$ in the conduction band for two different charge states. This cannot be accomplished by just the one 
 interaction parameters 
 of the Anderson model \cite{Anderson1961} which is sufficient near particle-hole symmetry as in the Kondo impurity.
 At least another independent parameter is required for screening.
 The 
Hartree-Fock mean-field phase diagram of the magnetic impurity with tunable screening interactions 
 \cite{Haldane1977_mixedval} gives a first order phase transition at the mixed-valence point. This while revealing is insufficient to the problem. 
 The problem was solved by the
  Wilson numerical 
 renormalization group (RG) \cite{PerakisVRuck}, which for parameters satisfying the Friedel theorem was discovered to
 have logarithmic low energy singularities similar to the singularities in the two channel Kondo problem and 
 the two Kondo impurity problem. It was further investigated analytically to derive the same singularities using Abelian bosonization 
 in \cite{Sire_V_R_G}. I show in this paper that these results may be expressed in terms of Majorana excitations. 
   
 The physical reason for the singularity and the associated Majorana at a mixed-valence may be explained with reference to Fig. (1) in 
 Ref. \cite{CMVcorrIns}
 where it is argued that an insulator in rare-earth compounds must with high probability in 
 parameter space have the magnetic ions be in the mixed-valence state rather than of the simple Kondo  state. In such compounds 
 the system changes from a fermi-liquid with one phase shift of the fermions when the correlated orbital is 
 above the chemical potential, i..e  the impurity has one particular valence,
 to a Fermi-liquid with a different phase shift when the parameters are such that the correlated orbital is  below the chemical potential where the impurity
  has the adjacent valence.  
 The two Fermi-liquids have different symmetries \cite{PerakisVRuck} so that one cannot pass continuously from one to the other as the parameters are varied through
 the mixed-valence condition. Achieving the condition in short-range interaction models requires tuning at least one parameter. 
 In a lattice of correlated orbitals, the singularity pins the chemical potential to the mixed valence  condition over a wide region of parameters and
 temperature \cite{Ruck_V}. 

This paper is organized as follows. First the experimental results are very briefly summarized.  In Sec. III, the starting Hamiltonian used in 
 \cite{PerakisVRuck} 
and \cite{Sire_V_R_G} 
is introduced. In \cite{Sire_V_R_G}, a specific bosonization procedure and rotations or canonical transformations are used to 
simplify the problem. An alternative canonical transformation is shown to give mathematically similar results. 
Unlike the two channel Kondo problem but like the two Kondo impurity problem, Abelian bosonization does not reduce
 the problem to quadrature. A strong coupling procedure introduced for the two Kondo impurity problem \cite{SVK}
  is used to show that the mixed valence problem also
 has super-symmetry, i.e. bosonic and fermionic ground states have identical energy, as do the excitations.  
 This appears necessary to show that the single particle excitations are Majoranas. 
 The approximate extension to the
lattice in terms of these variables is presented in Sec. IV.  It is shown that  when Bloch rules for insulators are satisfied, a band of Majorana
hybridizes with local resonances to present a gap in the spectra while another band of Majoranas remains free across the chemical potential. 
The approximation is to independently perform the transformations 
done in the single-impurity problem to every site in the periodic lattice. Some justification of this procedure is given. 
The application to the magneto-oscillations due to the gapless Majorana band, without linear coupling to low energy electromagnetic fields,
 is explained.
This is followed by a brief discussion of some important issues which need further work for a better understanding of the excitations.
Important experimental tests of the theory are suggested. Two principal predictions are extensive ground state entropy and the
absence of Zeeman splitting of the magneto-oscillations.

 \section{Experiments}
 
The most complete experimental evidence 
 for anomalous properties we are concerned with are available in 
 SmB$_6$ although similar properties are observed in YbB$_{12}$.  The symmetry of the band-structure in  SmB$_6$ 
 is such that it is also a topological insulator \cite{Dzero_Coleman} with convincing
 evidence for surface states \cite{Greene2016TopoSmB6}.
 SmB$_6$ was investigated and characterized to be a mixed-valence insulator long ago, in 1968 \cite{Geballe1971}. 
  Two groups have observed nearly similar magneto-oscillations in this compound. 
Arguments have been given by one experimental group \cite{Li2014SmB6} that the magneto-oscillations in SmB$_6$,
 claimed to be due to bulk excitations by the other \cite{Sebastian2015SmB6} may be due to surface states, while for YbB$_{12}$, there seems to be agreement
 among both groups \cite{ Li2018YbB12, Sebastian2018YbB12} for magneto-oscillations due to bulk excitations.
I regard the conclusion that the magneto-oscillations periodic in $1/H$, under discussion, in insulating SmB$_6$  are due to
bulk excitations as trustworthy for a few reasons.  They are accompanied 
  by "electronic" heat capacity in the insulator
with magnitude of about 5 mJ/mole $K^2$ at low temperatures. For comparison the specific heat of metallic Cu is 1/2 mJ/mole K$^2$.
 The amplitude of
 the  oscillations is similar to their diamagnetic susceptibility which is a bulk property. 
 The oscillations are also accompanied by low energy magnetic fluctuations
which were already in evidence in $\mu$SR measurements \cite{BiswasMusR}. The $\mu$SR relaxation rate becomes 
constant below about 4 K down to the lowest measured temperature of 19 mK. $\mu$SR relaxation is a probe for bulk excitations
in the limit of zero energy. A constant in temperature relaxation rate is itself a major anomaly.

It is also worth noting that the low temperature electronic specific heat of samples made by different groups \cite{PhelanSphtSmB6}, 
\cite{Sebastian2018YbB12} are remarkably different. This is further discussed in the last section of the paper.
  
\section{The mixed valence impurity}

I first recall here some features of Perakis et al.\cite{PerakisVRuck} and 
Sire et al. \cite{Sire_V_R_G}, which are essential for this work. I start with essentially the same
  Hamiltonian, written in the Wannier orbitals of the form used by Wilson \cite{KWilson} in 
  two different symmetries, the  fermions $h_{i\sigma}$ which hybridize with the local orbitals $f_{\sigma}$
  which have a large local repulsion parameters, and the screening fermions $s_{i\sigma}$ in a symmetry
  which is orthogonal to that of the hybridizing channels with a repulsive interaction between them.
  \be
  \label{H}
  H &=& H_0 + H_{kin}\\ 
H_0 &=& \epsilon_f (n_f -1/2) + U n_{f\uparrow} n_{f\downarrow} + t \sum_{\sigma} f_{\sigma}^+h_{0\sigma} + H.C. + V(n_f-1/2)(n_{s0}-1/2), \\
H_{kin} &=& \sum_{i,\sigma} t_h ~h^+_{i\sigma} h_{i+1 \sigma} + t_s s^+_{i\sigma} s_{i+1 \sigma} + H.C.
\ee
$i$ labels the sites of the one-dimensional Wilson chain. The parameter $V$ is absent in the usual Anderson model \cite{Anderson1961}. 
 It is assumed that both $U, V >> t$, 
the hybridization parameter $t$ for the local 
  orbital. Actually $U \to \infty$.     
The two lowest energy states in $H_0$ are multi-particle (excitonic) states,
\be
\label{states}
\zeta_{\sigma}^+|0> &=& |\sigma, 0>, ~{\text with ~ energy} ~E_{\zeta} = -\big((\epsilon_f/2 - V/4)^2 + t^2\big)^{1/2}, \\
\eta^+|0> &=& |0, 1>, ~{\text with ~energy} ~E_{\eta} = -(\epsilon_f/2 +V/4).
\ee
Here, the first entry in the bracket is the spin of the charge and spin of the impurity plus the hybridized charge of the neighboring sites
in the symmetry of the orbitals of the impurity.
The second is the charge of the screening channel. The charge in the screening channel is either 0 or 1 of either spin. This can be properly
counted by defining the operator
\be
\label{s}
s_0 = s_{0,\sigma}(1-n_{s, 0,-\sigma}) +  s_{0, -\sigma}(1-n_{0, s, \sigma}), ~~ n_{0, s, \sigma} \equiv s^+_{0, \sigma} s_{0, \sigma}.
\ee
  The constraint on the states is 
\be
\label{constraint}
 \sum_{\sigma} \zeta^+_{\sigma}\zeta_{\sigma} +  \eta^+\eta  =1.
\ee 
The two states $\zeta^+_{\sigma}|0>$ and $\eta^+|0>$ satisfy the essential requirement of the Friedel theorem. 

 The effective Hamiltonian in the space of these two states  obtained for $t/V << 1$ 
is \cite{Sire_V_R_G}, 
\be
\label{heff}
H_{eff} &=& H_{kin} +\frac{\epsilon}{2}\big(\sum_{\sigma} n_{\zeta, \sigma} - n_{\eta}\big) + \hat{t} \big( \sum_{\sigma} \zeta_{\sigma}^+ \eta s_0^+
h_{0 \sigma} + H.C. \big)\\ \nonumber
&+& J \zeta_{\sigma}^+ {\bf \sigma} \zeta_{\sigma} \cdot  h^+_0 {\bf \sigma} h_0
+ V_s(n_{\zeta}-1) (n_{\eta} -1).
\ee
In (\ref{heff}) $\epsilon = (E_{\zeta} - E_{\eta})$, and the parameters $J, V_s,$ and $\hat{t}$ are of similar magnitudes,
\be
J \approx \frac{2 t_h^2}{(\epsilon + V)},~~ V_s \approx  \frac{2 t_s^2}{(\epsilon + V)},~~ \hat{t} \approx  \frac{2 t_h t_s V}{(\epsilon^2 + V^2)}.
\ee

The direct hybridization in the initial Hamiltonian disappears in (\ref{heff}) in the $\zeta-\eta$ space  and  replaced by 
 multi-particle resonances coupling to kinetic energy (K.E.) which carry the effect farther. Secondly, the parameters 
 $\hat{t}, J, V_s$ are all marginal. The terms in kinetic energy coupling the local resonances to farther and farther Wilson orbitals
  are successively irrelevant to  
  $O(\Lambda^{n/2})$, where $\Lambda << 1$ is the RG expansion parameter.

\subsection{Bosonization and re-Fermionization}

One can represent the one-dimensional hybridizing fermions coupling to the impurity
by boson operators $\Phi_{h \sigma} (x)$ 
\be
\label{bosons}
 h_{\sigma}(x) = \frac{e^{-i \Phi_{h\sigma}(x)}}{\sqrt{2\pi a}}.
\ee
$\Phi_{h \sigma} (x)$  is governed by the appropriate Lagrangian given for example in standard works on bosonization \cite{Giamarchi}
One may now rotate the obtained Hamiltonian by the operator
$U = T_{\sigma}T_{-\sigma}$, 
\be
T_{\sigma} = e^{i(\zeta_{\sigma}^+\zeta_{\sigma} -1/2) \Phi_{h, \sigma}(0)},
\ee 
and note that $T_{\sigma}$ and $T_{-\sigma}$ commute. The result \cite{Sire_V_R_G} is the Hamiltonian
\be
\label{b-H-1}
H = H_{kin} + \epsilon \big( \sum_{\sigma} (\zeta_{\sigma}^+\zeta_{\sigma} -1/2) - \eta^+\eta \big) + 
\hat{t} \big(s_0^+ \eta \sum_{\sigma} \zeta_{\sigma}^+  + H.C.\big) + J \sum_{\sigma}\zeta_{\sigma}^+ \zeta_{-\sigma}.
\ee
There are also some additional longitudinal terms introduced (noted in \cite{Sire_V_R_G})
 which do not affect the results obtained here and serve mostly to renormalize $\epsilon$ and so only alter the condition for criticality.
 In \cite{Sire_V_R_G}, Eq. (\ref{b-H-1})  is used to calculate the free-energy and the response functions and to show the low energy
 singularities. 
 
The Hamitonian (\ref{b-H-1}) is equivalent in essential mathematical aspects to that for the two-impurity Kondo problem
 as treated in \cite{SVK}. To show this 
 the term proportional to $J$ is written, using the constraint (\ref{constraint}) and taking 
 $U \to \infty$, as
 \be
 J \sum_{\sigma}\zeta_{\sigma}^+ \zeta_{-\sigma} \to  J \sum_{\sigma}\zeta_{\sigma}^+ \eta \eta^+ \zeta_{-\sigma}.
 \ee
With this change, $H$ is equivalent to Eq. (4) of Ref.  \cite{SVK} on the two impurity problem (after the rotation to $\Phi_s =0$ in the latter), and the identification
of $\eta^+ \zeta_{\uparrow}$ with $S_1^+$, $\eta^+ \zeta_{\downarrow}$ with $S_2^+$; $\hat{t}$ with $J$ and $J$ with $K$. The
other coefficients $J_z, J_m, J_{mz}$ present in \cite{SVK} are absent here. They did not play any essential role in \cite{SVK}. One must also 
also identify the two channels $1$ and $2$ in \cite{SVK} with the channels $s$ and $h$ here. 
One could define 
\be
\label{J}
\tau^+_{\uparrow} = \eta^+ \zeta_{\uparrow},~~\tau^+_{\downarrow} = \eta^+ \zeta_{\downarrow},~~ \tau^z_{\uparrow} = \eta^+\eta - \zeta^+_{\uparrow} \zeta_{\uparrow},
\ee
etc.  ${\bf \tau}_{\uparrow}$, ${\bf \tau}_{\downarrow}$ are two non-commuting spinors, just as 
${\bf S}_1, {\bf S}_2$ in \cite{SVK}.

Just as the two impurity problem, the 
present problem is 
not reducible to quadrature by Abelian bosonization. Instead, a strong coupling scheme was developed which I follow below.

\subsubsection{A different Bosonization and re-fermionization} 

Before we proceed, it is important to note that $H$ in Eq. (\ref{b-H-1}) is also obtained in terms of somewhat different variables by a different
rotation operator. We may bosonize the screening channel fermions also by
\be 
s(x) = \frac{e^{-i \Phi_{s}(x)}}{\sqrt{2\pi a}}
\ee
and introduce charge and spin bosons in the $h$-channel by
\be
\Phi_{hc}(x) = \frac{1}{2}(\Phi_{h, \uparrow}(x) +\Phi_{h, \downarrow}(x)); ~~\Phi_{hs}(x) =
 \frac{1}{2}(\Phi_{h, \uparrow}(x) -\Phi_{h, \downarrow}(x)). 
\ee 
The term in (\ref{heff}) proportional to $\hat{t}$ becomes
\be
\hat{t} ~ \frac{e^{-i \Phi_{s}(x)}}{{2\pi a}} \eta \big((\zeta^+_{\uparrow} + \zeta^+_{\downarrow}) e^{i \Phi_{hc}(0)}\cos(\Phi_{hs}(0)) 
+ i (\zeta^+_{\uparrow} - \zeta^+_{\downarrow}) e^{i \Phi_{hc}(0)}\sin (\Phi_{hs}(0)) \big) + H.C.
\ee
A rotation $U' = e^{i (\zeta^+_{\uparrow} \zeta_{\uparrow} - \zeta^+_{\downarrow} \zeta_{\downarrow}) \Phi_{hs}(0)}$,
and $U = e^{i \eta^+\eta \Phi_{s}(0)}$ puts $\Phi_{hs}(0) \to 0$ eliminating the term with the difference of the spin-directions as well.
 One may now transform the Hamiltonian (\ref{heff}) to (\ref{b-H-1}), with $s$ replaced by the fermion obtained by the inverse
 transformation back to fermion of the operator $(\Phi_{hc})$. In both procedures only one linear coupling of the spin-directions
 is coupled to the fermions, the other is left free.
 
 If instead, we rotate and eliminate $(\Phi_s - \Phi_{hc})$ out of the Hamiltonian, keeping the spin-fluctuations $\Phi_{hs}$ 
 and fermionizing it, it is not possible to make further progress, primarily because of the exchange term proportional to $J$
 which now appears as a product of two spinors and a fermion operator. The basic physics for mixed-
 valent criticality is in the charge sector. In the two cases, where the procedure works, the effective fermion is the representation of 
 as a screening density dressed with the charge and spin-fluctuations of the hybridizing channel, or the density of the hybridizing channel
 dressed with the charge density of the screening channel and the spin-density of the hybridizing channel. The mysterious rotations,
 which are essential for the final simple results,
 may be thought of as imposing boundary conditions and thereby non-local renormalizations of the channel that is kept. These boundary conditions
 are the essential ingredients of the conformal field theory solutions of such impurity problems \cite{AL2}.

\subsection{Strong-coupling scheme, supersymmetry and Majoranas}

The strong coupling scheme starts with ${\hat{t}}, J >> t_h, t_s$, as appropriate in the renormalization group because the successive
couplings $t_h, t_s$ in Wilson shells are $O(\Lambda^{n/2})$ irrelevant, while ${\hat{t}}$ and $J$ are of $O(\Lambda)^0$. So, we 
first diagonalize only the terms involving the latter. In doing so, one notices that since they are linearly coupled to fermions,
$\tau_{\sigma}$ must also be expressed in terms of fermions. One runs then into the same problem as in \cite{SVK} 
that $\tau_{\uparrow}, 
\tau_{\downarrow}$ commute while their fermion representation $d_{\uparrow}, d_{\downarrow}$ anti-commute. This is remedied 
by an effective Jordan-Wigner procedure by which
\be
\tau^-_{\sigma} & \to & d_{\sigma}\big( 1 - (1-i) n_{- \sigma}\big), ~~ \tau^+_{\sigma}  \to  \big( 1 - (1+i) n_{- \sigma}\big)d^+_{\sigma}, \\ \nonumber
 \tau^z_{\sigma} & = & d^+_{\sigma}d_{\sigma} - 1/2 = n_{\sigma} -1/2.
\ee
The $d_{\sigma}$'s then obey the anti-commutation relations while the $\tau_{\sigma}$'s obey commutation relations. 
The term proportional to $J$ is not affected by this transformation. 
The local Hamiltonian in terms of the $d_{\sigma}$'s is now
\be
\label{H-b-2}
H_{loc} = \frac{J}{2} \big(d^+_{\uparrow}d_{\downarrow} + d^+_{\downarrow}d_{\uparrow}\big) + 
\frac{\hat{t}}{2} \Big(c d^+_{\uparrow}\big(1-(1-i)n_{\downarrow}\big) - c d_{\downarrow}\big(1-(1-i)n_{\uparrow}\big) + H.C. \Big).
\ee
Here $c$ is the annihilation operator for the local fermion which could be either of the choices made above. 
$H_{kin} + H_{loc}$ is the exact representation of the
mixed valence impurity problem.

One can diagonalize (\ref{H-b-2}) in the space of $n_0 = c^+_0c_0 = 0,1; n_{\uparrow} =0,1; n_{\downarrow} =0,1$. The eight states
 organize into two un-coupled spaces, four with even total number and four with odd total number of "particles". 
 The basis vectors for the even 
 sub-space are \\ $|0,0,0>; |0,1,1>; |1,1,0>; |1,0,1>$,\\
  and the odd sub-space are \\ $|1,0,0>, |1,1,1>, |0,0,1>, |0,1,0>$.\\
   The three numbers
 in the bracket are $n_0, n_{\uparrow}$ and $n_{\downarrow}$, respectively.
 The eigenvalues and the eigenvectors of the Hamiltonians in these sub-spaces are given in an Appendix A, where the
 following results are derived: \\
 A.  The eigenvalues in
 each subspace split in to the ground state $E_G$, a pair of degenerate excited states $E_1=E_2$ and a higher excited
 state $E_3$. The energies  $E_G$, $E_1$, $E_2$ and $E_3$ are identical  for the even and the odd sectors. \\
 B.  Let us call these
 states, respectively, 
 \\ $|G,e>, |G,o>, |1,e>, |1,o>, |2,e>, |2,o>, |3,e>, |3,o>$. \\ 
 The even and the odd sectors
 may be regarded as bosonic and fermionic respectively, the degeneracy of the two sectors represents
  a super-symmetry in the problem. \\
 C.  Let $r \equiv J/{\hat{t}}$. Then for  
 \be
 r \sqrt{r^2+2} = 1, ~i.e.~ J \approx 0.643~ {\hat{t}},
 \ee
 which will be considered the condition for mixed-valence criticality,
 \be
 \label{majcond}
 <G,e|c_0|G,o> = <G,o|c_0|G,e> = <G,e|c^+_0|G,o> = <G,o|c^+_0|G,e>.
 \ee
 This means that $c_0$ is a real (Majorana) fermion which may be written as 
 \be
 c_0 = c_0^+ = \frac{a_0 + a_0^+}{\sqrt{2}},
 \ee
 The conjugate Majorana
 \be
\bar{c}_0 = i \frac{a_0 - a_0^+}{\sqrt{2}}
\ee
is un-coupled to the local resonances. $a_0, a_0^+$ are canonical complex spin-less fermions. 
$c_0, \bar{c}_0$ also obey fermion anti-commutation
relations. \\
D.  The matrix elements between the ground state of the even sector and the excited states of the 
odd sector, and the 
ground state of the odd sector and the excited states of the even sector 
of $c_0$ have the same property (\ref{majcond}) as between the ground state. Similarly all such
matrix elements of $\bar{c}_0$ are zero. This means that coupling to higher states retains 
the reality of $c_0, \bar{c}_0$. The coupling between
the resonances at different sites through the propagating fermions $c_i$ is therefore absent.

It also follows that $ \bar{c}_0$, which as shown have no coupling to the degenerate ground states, have no coupling to the
excited states as well.

To complete the strong-coupling development, 
the next Wilson orbitals $c_1$ in the kinetic energy is coupled to  $c_0$ leading to new degenerate ground 
states, so that $c_1$ can be shown to be a Majorana also, and successively for all orbitals $c_n$
which are increasingly irrelevant to $O(\Lambda^{n/2})$. This is also done for the two-channel problem \cite{Emery-Kivelson92}
and the two-impurity problem \cite{SVK} and has its antecedent in the procedure due to Nozi\'eres and Blandin \cite{Nozieres_Blandin}.
In their conformal field theory solution of the two-Kondo impurity problem, Affleck and Ludwig \cite{AL2} showed that the symmetry of the
Majoranas in the problem have an Ising symmetry. This is true for the present problem and is different from the symmetry of the Majoranas in 
the two-channel 
problem \cite{AL1}. 
This is to be contrasted with the Majoranas introduced by Kitaev \cite{Kitaev2001} in a one-dimensional superconducting wire, which 
by construction have
$U(1)$ symmetry. 

Since only one linear combination of the local operators $(\tau_{\uparrow} + \tau_{\downarrow})$ is coupled to fermions, 
and the other $(\tau_{\uparrow} - \tau_{\downarrow})$ is left free,
there is ground state entropy
$\frac{1}{2} k_B \ln 2$ for the spin-1/2 mixed-valence impurity problem as there is in the two-channel Kondo impurity problem
and the two-impurity Kondo problem at criticality.

Just as in Ref. \cite{SVK}, one could perform a mean-field approximation on Eq. (\ref{H-b-2}) and obtain an effective Hamiltonian,
\be
H_K + k (d_{\uparrow}^+ d_{\downarrow} + H.C.) + i p (d_{\uparrow}^+ d^+_{\downarrow} + d_{\uparrow} d_{\downarrow}), 
\ee 
where $k$ and $p$ are parameters which depend self-consistently on expectation values such as $<c_0 d_{\sigma}>,  <c_0 d^+_{\sigma}>$.
At the critical condition $k=p$,  the 
 correlation functions and eigenvalues are
 are the same as from the strong coupling solution and the solution with the Wilson numerical RG. 
 The fact that mean-field theory works is as equally surprising in this problem as for the two impurity problem
 and makes one think that there is some more clever transformation with which the problem is exactly quadratic.
  We shall not pursue 
 this mean-field theory further here because it does not add to the discussion of the lattice problem.

 \section{Theory for the Mixed-Valence Lattice}
 
  In the problem of the heavy-fermion lattice, many particle physics is considered usually
 only at the level of the single site Kondo problem. 
The spatial periodicity of the resulting phase shifts of the light bands in the local symmetry of the 
local moment is then used to generate a band-structure with an energy scale of the Kondo temperature $T_K$, which is 
 the characteristic universal scale of change of phase shift with energy. This approach is remarkably successful in comparing low energy and low
temperature physics with experiments, except when the interactions between the
resonances due to their direct interactions with each other are comparable to $T_K$. A $1/N$ expansion in the number of channels of fermions and/or
number of orbital degree of freedom formally allows a justification. The physical reason for the success 
of the approximation is that due to the scale of parameters in the problem, with the 
Kondo temperature much less than the electronic band-width, the self-energy has negligible 
momentum independence compared to the energy dependence.  
This in turn
makes vertex renormalizations unimportant. 
A proof that this procedure works is that experimental Fermi-surfaces \cite{Lonzarich1988FS_UPt3}
of the heavy-fermions  have almost the same size and angular dependence as in one-particle calculations but with 
strongly renormalized mass and velocities. This follows from the theory for the fermi-liquid renormalization
when the self-energy satisfies the condition mentioned \cite{Varma_HF_phen}.
 
I follow the same approach in using the many-body resonances obtained above for the mixed-valence impurity
problem to the periodic array of magnetic ions with local correlated orbitals as well as itinerant fermions. 
 Especially in an insulator and the fact that the Majoranas are free would 
appear to make the procedure work as well for the metallic states. But ultimately, here as there,
 this may be proven satisfactory only if the experiments 
satisfy the predictions of the resulting 
theory.

Let us first consider the lattice problem in which the mixed-valence problem has been solved at each site independently.
The lattice Hamiltonian is
\be
\label{lattice}
H = H_K + G \sum_i \big(|e,i><o,i| + |o,i><e,i|\big)~ c_{i}.
\ee
$|e,i>$ and $|o,i>$ are the degenerate ground states for the bosonic and the fermionic local Hamiltonians noted above 
at sites $i$. The second term comes from Eq. (\ref{majcond}) at each site $i$. $H_K$ is the 
kinetic energy of the fermions in terms of the spin-less operators $c_i$ and ${\bar{c}_i}$. $c_{i}$ is the local Majorana at site $i$ which couples to them 
and $\bar{c}_{i}$  is the local Majorana which
does not. $G$ is the gap calculated at the strong coupling limit to be  $\hat{t}/(2 + r^2) \approx \hat{t}/\sqrt{2.43}$ but which
will have perturbative numerical renormalizations due to coupling to the excited local states.

The meaning of the operators $c_{i}, \bar{c}_i$ in terms of the fermions in the starting Hamiltonian represented by the operators $s_{i,\sigma}, 
s^+_{i,\sigma}; h_{i,\sigma}, h^+_{i,\sigma}$ is different in the two procedures described above. Strictly speaking, $c_i$ is the fermion representation
of the density operator in the one-dimensional chain terminating at the site $i$ in the screening and the hybridizing channel in the two
procedures respectively. 
We can think of them in the first procedure $c_i$ as the  sum over
the two spins of the $s_{i, \sigma}$-operator (so that it is a spin-less operator quantized in the ${x}$-direction), 
but the bosonization procedure dresses it with a multi-particle density fluctuations as well as the 
spin-fluctuations of the $h$- fermions. In the second procedure, as discussed above, $c_{i}$ is the sum over the two spins of the $h_{i,\sigma}$
operators as dressed by the density fluctuations of the $s$-channel and the spin-fluctuations of the $h$-channel. 
One is inclined to
suspect that the two procedures describe the same effective multi-particle resonances with fermion rules.
 In either case, this may be justified only
by the fact that the correct symmetries are retained. For some rationalization of the procedure,
recall the Toulouse \cite{Toulouse1970} bosonization and refermionization procedure of the Kondo problem.
 If the final fermion is regarded as the dressed
fermion of the hybridizing channel, the correct answers are obtained for the heavy-Fermion lattice.
 In any event, any further results are only  based on
objects with symmetry of fermions coupling to the local resonances as derived to get (\ref{lattice}).

\subsubsection{\bf Lattice Majoranas}

It is always possible to represent fermions on a lattice by Majornana fermions.
Looking at the form of the Hamiltonian (\ref{heff}) and subsequent developments, 
it is obvious that the Majoranas
must be written so that they are spin-less or equivalently as in (\ref{ferm}) point in the x-direction only. 
So we consider an effective free Hamiltonian of spin-less canonical fermions
 with kinetic energy 
\be
H_K = i t \sum_{i, n(i)} ( a_{i }^+a_{i+n} - a_{i+n}^+a_{i}).
 \ee
 $n(i)$ are neighbors of $i$. It is convenient as above to take the kinetic energy parameter to be imaginary. 
 On a bi-partite lattice, this is
 simply a phase shift of $\pi/2$ at alternate lattice points. Consider the Majorana operators
 \be
\label{ferm}
c_{i} = \frac{1}{\sqrt{2}}  (a_{i} + a_{i}^+),  \\
\bar{c}_{i} = i \frac{1}{\sqrt{2}} (a_{i} - a_{i}^+).
\ee
Then
 \be
 H_K  = i t \big(\sum_{i, n(i)}  c_{i} c_{i+n} +  \bar{c}_{i} \bar{c}_{i+n}\big),
 \ee
$n(i)$ denotes neighbors of a site $i$. (This is different from the Kitaev odd-parity superconducting model in which 
biquadratic terms off-diagonal in the conjugate Majoranas appear.)

 Note that the pairing operators in terms of the fermions  cancel out in $H_K$ although they are present in the two terms
 separately. Since the local operators couple only to $c_i$, we expect to find different spectra for the two terms. Then the two terms in $H_K$ each do
 have pairing terms. This effectively breaks $U(1)$ invariance but without any implication of superconductivity. 
 
 Let us diagonalize $H_K$ by transforming to momentum space. It is best to go back to the original variables to
 accomplish this. Let us call the two terms in $H_K$ as $H_{K1}$ and $H_{K2}$.
 \be
 H_{K1} &=& - 2 t \sum_{k, n} (\sin ({\bf k}. {\bf R_n})- \mu ) a_{{\bf k}}^+ a_{{\bf k}} 
 + 2 t \sum_{{\bf k}, n}\sin ({\bf k}. {\bf R_n}) (a_{{\bf k}}^+ a_{{\bf -k}}^+ - a_{{\bf k}} a_{{\bf -k}}); \\
  H_{K2} &=& - 2 t \sum_{k, n} (\sin ({\bf k}. {\bf R_n})- \mu ) a_{{\bf k}}^+ a_{{\bf k}} 
 - 2 t \sum_{\bf{k} n}\sin ({\bf k}. {\bf R_n}) (a_{{\bf k}}^+ a_{{\bf -k}}^+ - a_{{\bf k}} a_{{\bf -k}}).
  \ee
  Undoing the phase shift 
  introduced earlier, $- \sin ({\bf k}. {\bf R_n})$ changes to $\cos ({\bf k}. {\bf R_n})$
The sum of $H_{K1}$ and $H_{K2}$ has of-course no pairing term and gives simply the ordinary kinetic energy.
  
  Since $\bar{c}_i$ does not couple to the local resonances in the Hamiltonian (\ref{lattice}), $H_{K2}$ gives
   the dispersion of one-particle excitations,
  \be
  \label{Ev1}
  E_{2}({\bf k}) - \mu = 2t \sum_n \cos ({\bf k}. {\bf{R}}_n),
  \ee
  just that of free-fermions but the eigenvectors are the Majorana fermions $\bar{c}({\bf k})$.
  
  Taking into account the coupling of the $c_{i}$ Majoranas to the local  resonances in (\ref{lattice}) with the parameters $\hat{t},J$, the dispersion 
  of the bonding and anti-bonding orbitals is
  \be
  \label{Ev2}
  E_{1a}({\bf k}), E_{1b}({\bf k}) = \pm \sqrt{{G}^2 +  (E_{1}({\bf k})-\mu)^2}
  \ee
  with  $E_{1}({\bf k})=  E_{2}({\bf k})$. 
  
  $E_{2}({\bf k})$ is continuous across the chemical potential while $ E_{1a}({\bf k}), E_{1b}({\bf k})$ shows a maximum gap of $2G$. 
  Since a dissolution of the Majorana to canonical fermion can occur only
  if both $c$ and $\bar{c}$ are found at the same point with
  the same energy, the Majoranas $\bar{c}({\bf k})$ are protected. On the other hand, one expects for energies 
  larger than the gap $|G|$
  that the eigenvectors are of ordinary fermions. 
   $E_{1a, {\bf k}}, E_{1b, {\bf k}}$ is the generic nature of spectrum usually attributed to Kondo insulators. We might however expect that the 
   eigenvectors revert to that of canonical fermions at energies of order $G$ above the gaps (due to the "coherence factors").
   
     The compounds $SmB_6$ and $YbB_{12}$ are topological insulators due to their "inverted" band-structure \cite{Dzero_Coleman}. This does not
invalidate anything in relation to the exceptional bulk excitations derived here. It would of-course be worthwhile to do the above 
calculations for the real band-structure and also to see how the bulk Majoranas affect the surface states.

\subsection{Pinning of chemical potential at Mixed-valence}

Although achieving a mixed-valence impurity and associated criticality requires fine tuning of parameters, there is a pinning of the chemical
potential at mixed-valence for a finite range of parameters. The argument has already been given \cite{PerakisVRuck}, taking the lattice problem
as a periodic array of non-interacting impurities, and is merely repeated here for completeness. Let the 
 fraction of the ion with valence 1 be $<Q_1>$. Similarly, let $<Q_0>$ be the fraction of the
ions in valence 0. Both $<Q_0>$ and $<Q_1>$ are functions of $(\epsilon_f - \mu)$ and their sum must 
satisfy the condition for fixed charge $\bar{Q}$ for the lattice.
\be
<Q_0>+ <Q_1> = \bar{Q}
\ee
Near the mixed-valence condition on the chemical potential, the free-energy has a singularity so that 
\be
 <Q_1> \propto (\epsilon_f - \mu)^{1/\delta}.
 \ee
 $\delta \approx 2.7$ has been calculated by Wilson RG \cite{PerakisVRuck}. Then due to the singularity
 a self-consistent solution exists for a range of $\bar{Q}$ at the critical value. This procedure is similar to 
 finding the Pressure-Volume relation in a gas near its critical point. At finite temperature, there will be
 a cross-over to Fermi-liquid behavior, which has also been discussed \cite{PerakisVRuck}.

\section{Magneto-oscillations due to Majoranas}

The interesting idea that bulk Majorana fermions would produce magneto-oscillations 
was suggested by Baskaran \cite{Baskaran_Majorana} who used one of  the two superfluous free
  Majoranas in a representation of spin-ful conduction electrons in terms of four Majoranas in a theory for 
 pairing in time-reversal odd singlet superconductors in heavy fermion metals by Coleman et al. \cite{Coleman_Miranda}. 
 Quite evidently, SmB$_6$ 
 which is a topological insulator due to its band-structure \cite{Dzero_Coleman}, with evidence for surface states in agreement with such a 
 band-structure \cite{Greene2016TopoSmB6}, does not have a pairing gap of any variety.
However, the idea that Majoranas do have
de Haas van-Alphen oscillations appears to be sound.  To seek bulk Majoranas, one must 
study the nature of mixed-valence which is the common feature in all the insulators which have shown the oscillations.
 This is what has been done above. 
 
 A well defined theory for oscillations in 
 small gap insulators \cite{Knolle_Cooper} with an inverted band-structure at magnetic
 fields such that the  Zeeman energy is comparable to the gap is not applicable to Sm$B_6$ where oscillations exist 
 for magnetic field substantially smaller. There are also other more speculative ideas \cite{Chowdhury_Senthil} 
 also without cognizance
  of the mixed-valence nature of compounds showing oscillations in the insulating state.
 
 It is essential in the experiments in SmB$_6$ to have magneto-oscillations without a linear response to an electric field 
 that the system break gauge invariance.  Given this, it is worth recapitulating why Majoranas  produce de Haas - van Alphen oscillations;
  this has been explained by
 Erten et al. \cite{Erten_Coleman} particularly clearly. As shown above the Majoranas $\bar{c}({\bf k})$ crossing the chemical potential
 are a linear equal combination of a
 canonical particle and anti-particle operators. The linear coupling to a vector potential ${\bf A}$ changes their
energy to
 \be
 \label{A}
\frac{1}{2}\Big( E_2({\bf k} -\frac{e}{c}{\bf A}) + E_2({\bf k} +\frac{e}{c}{\bf A})\Big).
\ee
 The linear response to ${\bf A}$ (for example electrical conductivity) is then zero but effects quadratic in  ${\bf A}$
are present. The magneto-oscillations are effects due to change in free-energy which are functions of even powers of
 the magnetic field ${\bf B} = \nabla \times {\bf A}$. So, (\ref{A}) gives the same oscillations as the customary theory.
 The observed magneto-oscillations in SmB$_6$ match the
 fermi-surface of LaB$_6$ \cite{Sebastian2015SmB6}, i.e a band-structure similar to that in SmB$_6$ without the hybridization with
 local 4f-resonances. This is the result derived above. 
 The specific heat from Majorana excitations is also similar to that of
 fermions of similar density and mass.
 
 {\it Finite Temperature}:\\
 The results for for the mixed-valent impurity used here give the fixed point at $T\to 0$ and prove its validity
 by calculating the leading correction terms. It is not designed to give the quantitative values of the parameters and especially 
 not the crossovers at higher temperatures. This is important to know because experiments, especially in SmB$_6$ show 
 such cross-overs. More detailed calculations with study of spectra at various temperatures of the Wilson RG kind \cite{PerakisVRuck}
 may give such information. Application of the methods devised by Affleck and Ludwig \cite{AL1, AL2} might also be attempted.
 In Ref. \cite{Sire_V_R_G} such cross-overs are given as well as a dependence of the 
 chemical potential with temperature. The high temperature 
 crossover  occurs
 at the Kondo temperature of the mixed-valence problem which is of O($J^2 \rho$) where $\rho$ is the conduction 
 electron density
 of states as calculated in theories without considering screening. 
 This is the temperature below which one would begin to see
 Majoranas. The crossover to low temperatures is a more delicate matter; 
 it might be related to an exponentially lower
 scale related to ${\hat{t}}$. This and extension of such calculations to the lattice is left for future work. In the experiments there appear to be
 cross-overs near about 5 K marked by an increase in $C/T$ and muon relaxation rate
 and another cross-over below about 1 K where the amplitude of the oscillations and of $C/T$ increase very rapidly. The detailed temperature
 dependence of the latter needs to be studied in further experiments.
 
 \section{Predictions and further work indicated}
 
 The most interesting prediction of the theory, following from the approximation that many body effects due to scattering
of resonances at different sites are not important, concerns the entropy. As mentioned, half of the degrees of freedom of a mixed-valence 
impurity does not couple to the fermions if the effective spin is 1/2. In the present theory, this persists for each ion in the lattice. 
So a ground state
entropy of $O(1/2 ~R~ ln 2)/$mole is to be expected.
A confusing situation prevails in the available experimental results.
There are samples in which the measured magnetic entropy \cite{PhelanSphtSmB6} integrated to 300 K
is what is normally expected,
and there is no room for ground state entropy. However, in samples which have shown the magneto-oscillations, 
the entropy at 10 K obtained by integrating $C/T$ up to that temperature is 
 \cite{Sebastian2015SmB6} is  
 about a factor of 5 
smaller than in \cite{PhelanSphtSmB6}, and has not been measured at higher temperatures.
The integrated difference of magnetic  entropy at 10 K for the two measurements is not inconsiderable, 
about 0.3 Joules/mole/degree, but about 10 times less than 
$1/2$ R ln 2/mole. 
If the difference in $C/T$ 
persists 
in further measurements to room temperature,
 there is a case for ground state entropy of the magnitude suggested in the samples showing magneto-oscillations. 
(This also raises the question of the reason for such remarkable difference and 
the reason for the  large $C/T$ of about 30 mJ/mole/$K^2$ at low temperatures, well below its gap, 
 \cite{PhelanSphtSmB6}
 in an insulator.)
 
The approximation of non-interacting ions in the theory presented here
is bound to break down at low enough temperatures. 
Mixed valence $Sm^{2+}/Sm^{3+}$ ions have neither RKKY interactions nor 
double exchange interactions and neither do $Yb^{2+}/Yb^{3+}$ ions
\cite{cmvTmSe}.  But small higher order interactions are bound to be present. 
From the present experiments one can put an upper limit of 
about 1 K for their effects. Measurements of specific heat at lower temperatures are suggested.
 
The magneto-oscillation in SmB$_6$ are also accompanied by low energy magnetic fluctuations
in evidence in $\mu$SR measurements \cite{BiswasMusR}. 
The $\mu$SR relaxation rate becomes constant below about 2 K down to
the lowest measured temperature of 19 mK. 
Magnetic fluctuations which would give a temperature independent 
relaxation rate are derived for the single mixed valent impurity due to 
the $1/\tau$ magnetic correlations in imaginary
time, (or equivalently $\omega/T$ scaling), derived in in Ref. \cite{Sire_V_R_G} for the mixed-valence problem.
 These fluctuations need further investigation and direct measurements by neutron scattering. 
 
A test of the theory is also that the Zeeman splitting of the magneto-oscillations  due to a magnetic field 
should be absent as the Majoranas derived here are spin-less. 
This is also true of the gapped excitations - so their magnetic character close to the gap should 
show absence of linear effects in a field.

An obvious prediction is that the magneto-oscillations must be present both in the specific heat and thermal conductivity.

The pair-bilinears in the fermions representation of the diagonal representation in terms of Majorana bands 
may be detectable in tunneling experiments. 

Another results that follows from the present work concerns other materials in which Majoranas 
may be the proper description of low energy physics.
These may be varieties of spin-liquids and the so-called Kitaev compounds or 
those resembling them. Magneto-oscillations in them should be
looked for. \\

{\bf Acknowledgements}:  I wish to thank Suchitra Sebastian for a detailed discussion of the experimental results. 
As noted  in the paper, the foundation of the physics  in this paper 
were laid quite some time ago in the work with
Thierry Giamarchi, Ilias Perakis, Andrei Ruckenstein and Clement Sire, all of us then at Bell laboratories.
  I also wish to thank Piers Coleman for a recent discussion on the work of Erten et al.
and Srinivas Raghu for a general discussion on such problems.
This work was done as a "Recalled Professor" at Berkeley. I wish to thank James Analytis, Robert Birgeneau
and Joel Moore for their hospitality.

\section{Appendix}
{\it Solution of the local Hamiltonian for the single Mixed-valence impurity}\\

The Hamiltonian $H_{loc}$ of (\ref{H-b-2}) in the even particle basis \\ 
$|0,0,0>, |0,1,1>, |1,1,0>, |1,0,1>$ is \\
 \be 
 \label{Heven}
 H_{even}=
  \left(\begin{array}{cccc} 0 & 0 & -\hat{t}/2 & 0 \\0 & 0& -i \hat{t}/2 & 0 \\-\hat{t}/2 & i \hat{t}/2 & 0 & J/2 \\ 0 & 0 & J/2 & 0
 \end{array}\right)
 \ee 
  The ground has energy $E_0$ and eigenvector 
   \be 
   \label{g-even}
   E_{0e} &=& - \hat{t}/2 \sqrt{2 + r^2}; ~~r = J/\hat{t}\\  
   |0e> &=& u (|0,0,0> - i |0,1,1>) - v |1,1,0> + w |1,0,1>.
   \ee
   $$ u = \frac{1}{\sqrt{2(2+r^2)}},~~v = \frac{r}{\sqrt{2(2+r^2)}},~~w = \frac{1}{\sqrt{2}}.$$
   
   The next two states are degenerate with zero energy. Their eigenvectors after mutual orthogonalization are
   \be
   \label{ex1-even}
   E_{1e} &=& E_{2e} = 0; \\
    |1e> &=& \frac{1}{\sqrt{2}} (-i |0,0,0>  +  |0,1,1>); \\
     |2e> &=&  \Big(- \frac{r}{\sqrt{(1+r^2}} -i \frac{1}{2} \lambda) |0,0,0>    + \lambda  \frac{1}{\sqrt{2}}  |0,1,1>  +\frac{1}{\sqrt{1+r^2}} |101>\Big).\\
        \ee
        $\lambda$ is determined to orthogonalize the two degenerate eigenvectors and has the value
        $$\lambda =- i \frac{r}{\sqrt{1+r^2}}.$$
       
       The highest state has energy and eigenvectors
       \be
       \label{ex2-even}
       E_{3e} &=& (J/2) \sqrt{2 + r^2}; \\
       |3e> &=& u (|0,0,0> - i |0,1,1>) + v |1,1,0> + w |1,0,1>.
       \ee
       
       {\it Odd charge sector}
       
 The Hamiltonian $H_{loc}$ of (\ref{H-b-2}) in the odd number particle sector, i.e. in the basis 
\be
\label{basis-odd}
|1,0,0>, |1,1,1>, |0,0,1>, |0,1,0>
\ee
is
  \be 
  \label{Hodd}
 H_{odd}=
  \left(\begin{array}{cccc}0 & 0 & 0 & \hat{t}/2 \\0 & 0 & 0 & -i \hat{t}/2 \\0 & 0 &0& j/2 \\ \hat{t}/2 & i \hat{t}/2 & J/2 & 0
 \end{array}\right)
 \ee  
   The energy levels are identical to the even sector. The even sector is bosonic, the odd is fermionic.
   So there is  supersymmetry.: 
   
   The ground has energy $E_0$ and eigenvector 
   \be
   \label{g-odd}
   E_{0o} = - (\hat{t}/2) \sqrt{2 + r^2};~~ |0o> = -u (|1,0,0> - i |1,1,1>) - v |0,0,1> + w |0,1,0>.
   \ee
      
   The next two are degenerate with zero energy. Their eigenvectors after mutual orthogonalization are
   \be
   \label{ex1-odd}
   E_{1o} &=& E_{2o} = 0;  |1o> = -\frac{r}{\sqrt{1+r^2}}  |1,0,0>  + 1/\sqrt{1+r^2}|0,0,1> \\.
       |2o> &=& (- \lambda \frac{r}{\sqrt{1+r^2}}  -i /\sqrt{2}) |1,0,0> + 1/\sqrt{2} |1,1,1> +
        \lambda  \frac{1}{\sqrt{1+r^2}}|0,0,1>.
        \ee
    $\lambda$ is determined from the orthogonalization of the two degenerate states to be the same
    as earlier,
    $$ \lambda = -i \frac{r}{\sqrt{(1+r^2)}}.$$ 
    The highest energy state is
    \be
    \label{ex2odd}
      E_{3o} = (\hat{t}/2){\sqrt{2 + r^2}};   |3o> = u (|0,0,0> - i |0,1,1>) + v |1,1,0> + w |1,0,1>.
      \ee
       
   The important result is that (just as in the two impurity problem) if
       $$u^2 = 2vw,$$
       i.e.
       $$J\approx 0.634 \hat{t}$$  
       \be
       \label{majorana}
        < 0e|c_0| 0o> = < 0o|c_0| 0e>.
        \ee
       As discussed in the main text, this means that the fermion operator $c_0$ is purely real
       and therefore a Majorana. 
             
       The matrix element between the ground state and the excited
       state have exactly the same property, i.e
       \be
       \label{ex-maj}
        < 0e|c_{0p}| \alpha, o> &=&  < \alpha,o|c_{0p}| 0e>, for ~all~~ \alpha = 1,2,3.\\
         < 0o|c_{0p}| \alpha, e> &=&  < \alpha,e|c_{0p}| 0o>, for ~all~~ \alpha = 1,2,3 ~also.
         \ee
       and that $c_{ol}$ has zero matrix elements.
       
In the above, the diagonal terms which produce difference in the energy of the states were put to zero.
The same results are obtained when they are included but with a more complicated condition and coefficients.
The mixed-valence condition


\begin{thebibliography}{46}%
\makeatletter
\providecommand \@ifxundefined [1]{%
 \@ifx{#1\undefined}
}%
\providecommand \@ifnum [1]{%
 \ifnum #1\expandafter \@firstoftwo
 \else \expandafter \@secondoftwo
 \fi
}%
\providecommand \@ifx [1]{%
 \ifx #1\expandafter \@firstoftwo
 \else \expandafter \@secondoftwo
 \fi
}%
\providecommand \natexlab [1]{#1}%
\providecommand \enquote  [1]{``#1''}%
\providecommand \bibnamefont  [1]{#1}%
\providecommand \bibfnamefont [1]{#1}%
\providecommand \citenamefont [1]{#1}%
\providecommand \href@noop [0]{\@secondoftwo}%
\providecommand \href [0]{\begingroup \@sanitize@url \@href}%
\providecommand \@href[1]{\@@startlink{#1}\@@href}%
\providecommand \@@href[1]{\endgroup#1\@@endlink}%
\providecommand \@sanitize@url [0]{\catcode `\\12\catcode `\$12\catcode
  `\&12\catcode `\#12\catcode `\^12\catcode `\_12\catcode `\%12\relax}%
\providecommand \@@startlink[1]{}%
\providecommand \@@endlink[0]{}%
\providecommand \url  [0]{\begingroup\@sanitize@url \@url }%
\providecommand \@url [1]{\endgroup\@href {#1}{\urlprefix }}%
\providecommand \urlprefix  [0]{URL }%
\providecommand \Eprint [0]{\href }%
\providecommand \doibase [0]{http://dx.doi.org/}%
\providecommand \selectlanguage [0]{\@gobble}%
\providecommand \bibinfo  [0]{\@secondoftwo}%
\providecommand \bibfield  [0]{\@secondoftwo}%
\providecommand \translation [1]{[#1]}%
\providecommand \BibitemOpen [0]{}%
\providecommand \bibitemStop [0]{}%
\providecommand \bibitemNoStop [0]{.\EOS\space}%
\providecommand \EOS [0]{\spacefactor3000\relax}%
\providecommand \BibitemShut  [1]{\csname bibitem#1\endcsname}%
\let\auto@bib@innerbib\@empty
\bibitem [{\citenamefont {Anderson}(1961)}]{Anderson1961}%
  \BibitemOpen
  \bibfield  {author} {\bibinfo {author} {\bibfnamefont {P.~W.}\ \bibnamefont
  {Anderson}},\ }\href {\doibase 10.1103/PhysRev.124.41} {\bibfield  {journal}
  {\bibinfo  {journal} {Phys. Rev.}\ }\textbf {\bibinfo {volume} {124}},\
  \bibinfo {pages} {41} (\bibinfo {year} {1961})}\BibitemShut {NoStop}%
\bibitem [{\citenamefont {Varma}\ and\ \citenamefont {Yafet}(1976)}]{V_Yafet1}%
  \BibitemOpen
  \bibfield  {author} {\bibinfo {author} {\bibfnamefont {C.~M.}\ \bibnamefont
  {Varma}}\ and\ \bibinfo {author} {\bibfnamefont {Y.}~\bibnamefont {Yafet}},\
  }\href {\doibase 10.1103/PhysRevB.13.2950} {\bibfield  {journal} {\bibinfo
  {journal} {Phys. Rev. B}\ }\textbf {\bibinfo {volume} {13}},\ \bibinfo
  {pages} {2950} (\bibinfo {year} {1976})}\BibitemShut {NoStop}%
\bibitem [{\citenamefont {Varma}(1976)}]{V_RMP_MV}%
  \BibitemOpen
  \bibfield  {author} {\bibinfo {author} {\bibfnamefont {C.~M.}\ \bibnamefont
  {Varma}},\ }\href {\doibase 10.1103/RevModPhys.48.219} {\bibfield  {journal}
  {\bibinfo  {journal} {Rev. Mod. Phys.}\ }\textbf {\bibinfo {volume} {48}},\
  \bibinfo {pages} {219} (\bibinfo {year} {1976})}\BibitemShut {NoStop}%
\bibitem [{\citenamefont {Razafimandimby}\ \emph {et~al.}(1984)\citenamefont
  {Razafimandimby}, \citenamefont {Fulde},\ and\ \citenamefont
  {Keller}}]{Fulde_HF}%
  \BibitemOpen
  \bibfield  {author} {\bibinfo {author} {\bibfnamefont {H.}~\bibnamefont
  {Razafimandimby}}, \bibinfo {author} {\bibfnamefont {P.}~\bibnamefont
  {Fulde}}, \ and\ \bibinfo {author} {\bibfnamefont {J.}~\bibnamefont
  {Keller}},\ }\href {\doibase 10.1007/BF01388062} {\bibfield  {journal}
  {\bibinfo  {journal} {Zeitschrift fur Physik B}\ }\textbf {\bibinfo {volume}
  {54}},\ \bibinfo {pages} {111} (\bibinfo {year} {1984})}\BibitemShut
  {NoStop}%
\bibitem [{\citenamefont {Auerbach}\ and\ \citenamefont
  {Levin}(1986)}]{AuerbachLevin}%
  \BibitemOpen
  \bibfield  {author} {\bibinfo {author} {\bibfnamefont {A.}~\bibnamefont
  {Auerbach}}\ and\ \bibinfo {author} {\bibfnamefont {K.}~\bibnamefont
  {Levin}},\ }\href {\doibase 10.1103/PhysRevLett.57.877} {\bibfield  {journal}
  {\bibinfo  {journal} {Phys. Rev. Lett.}\ }\textbf {\bibinfo {volume} {57}},\
  \bibinfo {pages} {877} (\bibinfo {year} {1986})}\BibitemShut {NoStop}%
\bibitem [{\citenamefont {Millis}\ and\ \citenamefont
  {Lee}(1987)}]{MillisLeeHF}%
  \BibitemOpen
  \bibfield  {author} {\bibinfo {author} {\bibfnamefont {A.~J.}\ \bibnamefont
  {Millis}}\ and\ \bibinfo {author} {\bibfnamefont {P.~A.}\ \bibnamefont
  {Lee}},\ }\href {\doibase 10.1103/PhysRevB.35.3394} {\bibfield  {journal}
  {\bibinfo  {journal} {Phys. Rev. B}\ }\textbf {\bibinfo {volume} {35}},\
  \bibinfo {pages} {3394} (\bibinfo {year} {1987})}\BibitemShut {NoStop}%
\bibitem [{\citenamefont {Varma}\ \emph {et~al.}(1986)\citenamefont {Varma},
  \citenamefont {Weber},\ and\ \citenamefont {Randall}}]{V_Weber_Randall}%
  \BibitemOpen
  \bibfield  {author} {\bibinfo {author} {\bibfnamefont {C.~M.}\ \bibnamefont
  {Varma}}, \bibinfo {author} {\bibfnamefont {W.}~\bibnamefont {Weber}}, \ and\
  \bibinfo {author} {\bibfnamefont {L.~J.}\ \bibnamefont {Randall}},\ }\href
  {\doibase 10.1103/PhysRevB.33.1015} {\bibfield  {journal} {\bibinfo
  {journal} {Phys. Rev. B}\ }\textbf {\bibinfo {volume} {33}},\ \bibinfo
  {pages} {1015} (\bibinfo {year} {1986})}\BibitemShut {NoStop}%
\bibitem [{\citenamefont {Rice}\ and\ \citenamefont
  {Ueda}(1985)}]{Rice_Ueda_Gutzwill}%
  \BibitemOpen
  \bibfield  {author} {\bibinfo {author} {\bibfnamefont {T.~M.}\ \bibnamefont
  {Rice}}\ and\ \bibinfo {author} {\bibfnamefont {K.}~\bibnamefont {Ueda}},\
  }\href {\doibase 10.1103/PhysRevLett.55.995} {\bibfield  {journal} {\bibinfo
  {journal} {Phys. Rev. Lett.}\ }\textbf {\bibinfo {volume} {55}},\ \bibinfo
  {pages} {995} (\bibinfo {year} {1985})}\BibitemShut {NoStop}%
\bibitem [{\citenamefont {Georges}\ \emph {et~al.}(1996)\citenamefont
  {Georges}, \citenamefont {Kotliar}, \citenamefont {Krauth},\ and\
  \citenamefont {Rozenberg}}]{DMFT_RMP1996}%
  \BibitemOpen
  \bibfield  {author} {\bibinfo {author} {\bibfnamefont {A.}~\bibnamefont
  {Georges}}, \bibinfo {author} {\bibfnamefont {G.}~\bibnamefont {Kotliar}},
  \bibinfo {author} {\bibfnamefont {W.}~\bibnamefont {Krauth}}, \ and\ \bibinfo
  {author} {\bibfnamefont {M.~J.}\ \bibnamefont {Rozenberg}},\ }\href {\doibase
  10.1103/RevModPhys.68.13} {\bibfield  {journal} {\bibinfo  {journal} {Rev.
  Mod. Phys.}\ }\textbf {\bibinfo {volume} {68}},\ \bibinfo {pages} {13}
  (\bibinfo {year} {1996})}\BibitemShut {NoStop}%
\bibitem [{\citenamefont {Varma}(1985)}]{Varma_HF_phen}%
  \BibitemOpen
  \bibfield  {author} {\bibinfo {author} {\bibfnamefont {C.~M.}\ \bibnamefont
  {Varma}},\ }\href {\doibase 10.1103/PhysRevLett.55.2723} {\bibfield
  {journal} {\bibinfo  {journal} {Phys. Rev. Lett.}\ }\textbf {\bibinfo
  {volume} {55}},\ \bibinfo {pages} {2723} (\bibinfo {year}
  {1985})}\BibitemShut {NoStop}%
\bibitem [{\citenamefont {Tan}\ \emph {et~al.}(2015)\citenamefont {Tan},
  \citenamefont {Hsu}, \citenamefont {Zeng}, \citenamefont {Hatnean},
  \citenamefont {Harrison}, \citenamefont {Zhu}, \citenamefont {Hartstein},
  \citenamefont {Kiourlappou}, \citenamefont {Srivastava}, \citenamefont
  {Johannes}, \citenamefont {Murphy}, \citenamefont {Park}, \citenamefont
  {Balicas}, \citenamefont {Lonzarich}, \citenamefont {Balakrishnan},\ and\
  \citenamefont {Sebastian}}]{Sebastian2015SmB6}%
  \BibitemOpen
  \bibfield  {author} {\bibinfo {author} {\bibfnamefont {B.~S.}\ \bibnamefont
  {Tan}}, \bibinfo {author} {\bibfnamefont {Y.-T.}\ \bibnamefont {Hsu}},
  \bibinfo {author} {\bibfnamefont {B.}~\bibnamefont {Zeng}}, \bibinfo {author}
  {\bibfnamefont {M.~C.}\ \bibnamefont {Hatnean}}, \bibinfo {author}
  {\bibfnamefont {N.}~\bibnamefont {Harrison}}, \bibinfo {author}
  {\bibfnamefont {Z.}~\bibnamefont {Zhu}}, \bibinfo {author} {\bibfnamefont
  {M.}~\bibnamefont {Hartstein}}, \bibinfo {author} {\bibfnamefont
  {M.}~\bibnamefont {Kiourlappou}}, \bibinfo {author} {\bibfnamefont
  {A.}~\bibnamefont {Srivastava}}, \bibinfo {author} {\bibfnamefont {M.~D.}\
  \bibnamefont {Johannes}}, \bibinfo {author} {\bibfnamefont {T.~P.}\
  \bibnamefont {Murphy}}, \bibinfo {author} {\bibfnamefont {J.-H.}\
  \bibnamefont {Park}}, \bibinfo {author} {\bibfnamefont {L.}~\bibnamefont
  {Balicas}}, \bibinfo {author} {\bibfnamefont {G.~G.}\ \bibnamefont
  {Lonzarich}}, \bibinfo {author} {\bibfnamefont {G.}~\bibnamefont
  {Balakrishnan}}, \ and\ \bibinfo {author} {\bibfnamefont {S.~E.}\
  \bibnamefont {Sebastian}},\ }\href {\doibase 10.1126/science.aaa7974}
  {\bibfield  {journal} {\bibinfo  {journal} {Science}\ }\textbf {\bibinfo
  {volume} {349}},\ \bibinfo {pages} {287} (\bibinfo {year} {2015})},\ \Eprint
  {http://arxiv.org/abs/https://science.sciencemag.org/content/349/6245/287.full.pdf}
  {https://science.sciencemag.org/content/349/6245/287.full.pdf} \BibitemShut
  {NoStop}%
\bibitem [{\citenamefont {Xiang}\ \emph {et~al.}(2018)\citenamefont {Xiang},
  \citenamefont {Kasahara}, \citenamefont {Asaba}, \citenamefont {Lawson},
  \citenamefont {Tinsman}, \citenamefont {Chen}, \citenamefont {Sugimoto},
  \citenamefont {Kawaguchi}, \citenamefont {Sato}, \citenamefont {Li},
  \citenamefont {Yao}, \citenamefont {Chen}, \citenamefont {Iga}, \citenamefont
  {Singleton}, \citenamefont {Matsuda},\ and\ \citenamefont
  {Li}}]{Li2018YbB12}%
  \BibitemOpen
  \bibfield  {author} {\bibinfo {author} {\bibfnamefont {Z.}~\bibnamefont
  {Xiang}}, \bibinfo {author} {\bibfnamefont {Y.}~\bibnamefont {Kasahara}},
  \bibinfo {author} {\bibfnamefont {T.}~\bibnamefont {Asaba}}, \bibinfo
  {author} {\bibfnamefont {B.}~\bibnamefont {Lawson}}, \bibinfo {author}
  {\bibfnamefont {C.}~\bibnamefont {Tinsman}}, \bibinfo {author} {\bibfnamefont
  {L.}~\bibnamefont {Chen}}, \bibinfo {author} {\bibfnamefont {K.}~\bibnamefont
  {Sugimoto}}, \bibinfo {author} {\bibfnamefont {S.}~\bibnamefont {Kawaguchi}},
  \bibinfo {author} {\bibfnamefont {Y.}~\bibnamefont {Sato}}, \bibinfo {author}
  {\bibfnamefont {G.}~\bibnamefont {Li}}, \bibinfo {author} {\bibfnamefont
  {S.}~\bibnamefont {Yao}}, \bibinfo {author} {\bibfnamefont {Y.~L.}\
  \bibnamefont {Chen}}, \bibinfo {author} {\bibfnamefont {F.}~\bibnamefont
  {Iga}}, \bibinfo {author} {\bibfnamefont {J.}~\bibnamefont {Singleton}},
  \bibinfo {author} {\bibfnamefont {Y.}~\bibnamefont {Matsuda}}, \ and\
  \bibinfo {author} {\bibfnamefont {L.}~\bibnamefont {Li}},\ }\href {\doibase
  10.1126/science.aap9607} {\bibfield  {journal} {\bibinfo  {journal}
  {Science}\ }\textbf {\bibinfo {volume} {362}},\ \bibinfo {pages} {65}
  (\bibinfo {year} {2018})},\ \Eprint
  {http://arxiv.org/abs/https://science.sciencemag.org/content/362/6410/65.full.pdf}
  {https://science.sciencemag.org/content/362/6410/65.full.pdf} \BibitemShut
  {NoStop}%
\bibitem [{\citenamefont {Liu}\ \emph {et~al.}(2018)\citenamefont {Liu},
  \citenamefont {Hartstein}, \citenamefont {Wallace},\ and\ \citenamefont
  {et~al.}}]{Sebastian2018YbB12}%
  \BibitemOpen
  \bibfield  {author} {\bibinfo {author} {\bibfnamefont {H.}~\bibnamefont
  {Liu}}, \bibinfo {author} {\bibfnamefont {M.}~\bibnamefont {Hartstein}},
  \bibinfo {author} {\bibfnamefont {G.}~\bibnamefont {Wallace}}, \ and\
  \bibinfo {author} {\bibnamefont {et~al.}},\ }\href {\doibase
  10.1088/1361-648X/aaa522} {\bibfield  {journal} {\bibinfo  {journal} {J Phys
  Condens Matter}\ }\textbf {\bibinfo {volume} {30}} (\bibinfo {year} {2018}),\
  10.1088/1361-648X/aaa522}\BibitemShut {NoStop}%
\bibitem [{\citenamefont {Hartstein}\ and\ \citenamefont
  {et~al.}(2020)}]{SebastianSmB62020}%
  \BibitemOpen
  \bibfield  {author} {\bibinfo {author} {\bibfnamefont {M.}~\bibnamefont
  {Hartstein}}\ and\ \bibinfo {author} {\bibnamefont {et~al.}},\ }\href@noop {}
  {\bibfield  {journal} {\bibinfo  {journal} {arXiv:2007.01453}\ } (\bibinfo
  {year} {2020})}\BibitemShut {NoStop}%
\bibitem [{\citenamefont {Perakis}\ \emph {et~al.}(1993)\citenamefont
  {Perakis}, \citenamefont {Varma},\ and\ \citenamefont
  {Ruckenstein}}]{PerakisVRuck}%
  \BibitemOpen
  \bibfield  {author} {\bibinfo {author} {\bibfnamefont {I.~E.}\ \bibnamefont
  {Perakis}}, \bibinfo {author} {\bibfnamefont {C.~M.}\ \bibnamefont {Varma}},
  \ and\ \bibinfo {author} {\bibfnamefont {A.~E.}\ \bibnamefont
  {Ruckenstein}},\ }\href {\doibase 10.1103/PhysRevLett.70.3467} {\bibfield
  {journal} {\bibinfo  {journal} {Phys. Rev. Lett.}\ }\textbf {\bibinfo
  {volume} {70}},\ \bibinfo {pages} {3467} (\bibinfo {year}
  {1993})}\BibitemShut {NoStop}%
\bibitem [{\citenamefont {Sire}\ \emph {et~al.}(1994)\citenamefont {Sire},
  \citenamefont {Varma}, \citenamefont {Ruckenstein},\ and\ \citenamefont
  {Giamarchi}}]{Sire_V_R_G}%
  \BibitemOpen
  \bibfield  {author} {\bibinfo {author} {\bibfnamefont {C.}~\bibnamefont
  {Sire}}, \bibinfo {author} {\bibfnamefont {C.~M.}\ \bibnamefont {Varma}},
  \bibinfo {author} {\bibfnamefont {A.~E.}\ \bibnamefont {Ruckenstein}}, \ and\
  \bibinfo {author} {\bibfnamefont {T.}~\bibnamefont {Giamarchi}},\ }\href
  {\doibase 10.1103/PhysRevLett.72.2478} {\bibfield  {journal} {\bibinfo
  {journal} {Phys. Rev. Lett.}\ }\textbf {\bibinfo {volume} {72}},\ \bibinfo
  {pages} {2478} (\bibinfo {year} {1994})}\BibitemShut {NoStop}%
\bibitem [{\citenamefont {Nozi\'eres}(1974)}]{Nozieres_LFL}%
  \BibitemOpen
  \bibfield  {author} {\bibinfo {author} {\bibfnamefont {P.}~\bibnamefont
  {Nozi\'eres}},\ }\href {https://doi.org/10.1007/BF00654541} {\bibfield
  {journal} {\bibinfo  {journal} {Journal of Low Temperature Physics}\ }\textbf
  {\bibinfo {volume} {17}},\ \bibinfo {pages} {31} (\bibinfo {year}
  {1974})}\BibitemShut {NoStop}%
\bibitem [{\citenamefont {Kitaev}(2011)}]{Kitaev2001}%
  \BibitemOpen
  \bibfield  {author} {\bibinfo {author} {\bibfnamefont {A.}~\bibnamefont
  {Kitaev}},\ }\href@noop {} {\bibfield  {journal} {\bibinfo  {journal}
  {Physics-Uspekhi}\ }\textbf {\bibinfo {volume} {44}},\ \bibinfo {pages} {131}
  (\bibinfo {year} {2011})}\BibitemShut {NoStop}%
\bibitem [{\citenamefont {Balatsky}\ and\ \citenamefont
  {Abrahams}(1992)}]{Balatsky_Abrahams}%
  \BibitemOpen
  \bibfield  {author} {\bibinfo {author} {\bibfnamefont {A.}~\bibnamefont
  {Balatsky}}\ and\ \bibinfo {author} {\bibfnamefont {E.}~\bibnamefont
  {Abrahams}},\ }\href {\doibase 10.1103/PhysRevB.45.13125} {\bibfield
  {journal} {\bibinfo  {journal} {Phys. Rev. B}\ }\textbf {\bibinfo {volume}
  {45}},\ \bibinfo {pages} {13125} (\bibinfo {year} {1992})}\BibitemShut
  {NoStop}%
\bibitem [{\citenamefont {Coleman}\ \emph {et~al.}(1994)\citenamefont
  {Coleman}, \citenamefont {Miranda},\ and\ \citenamefont
  {Tsvelik}}]{Coleman_Miranda}%
  \BibitemOpen
  \bibfield  {author} {\bibinfo {author} {\bibfnamefont {P.}~\bibnamefont
  {Coleman}}, \bibinfo {author} {\bibfnamefont {E.}~\bibnamefont {Miranda}}, \
  and\ \bibinfo {author} {\bibfnamefont {A.}~\bibnamefont {Tsvelik}},\ }\href
  {\doibase 10.1103/PhysRevB.49.8955} {\bibfield  {journal} {\bibinfo
  {journal} {Phys. Rev. B}\ }\textbf {\bibinfo {volume} {49}},\ \bibinfo
  {pages} {8955} (\bibinfo {year} {1994})}\BibitemShut {NoStop}%
\bibitem [{\citenamefont {Baskaran}(2015)}]{Baskaran_Majorana}%
  \BibitemOpen
  \bibfield  {author} {\bibinfo {author} {\bibfnamefont {G.}~\bibnamefont
  {Baskaran}},\ }\href@noop {} {\bibfield  {journal} {\bibinfo  {journal}
  {arXiv:1507.03477}\ } (\bibinfo {year} {2015})}\BibitemShut {NoStop}%
\bibitem [{\citenamefont {Erten}\ \emph {et~al.}(2017)\citenamefont {Erten},
  \citenamefont {Chang}, \citenamefont {Coleman},\ and\ \citenamefont
  {Tsvelik}}]{Erten_Coleman}%
  \BibitemOpen
  \bibfield  {author} {\bibinfo {author} {\bibfnamefont {O.}~\bibnamefont
  {Erten}}, \bibinfo {author} {\bibfnamefont {P.-Y.}\ \bibnamefont {Chang}},
  \bibinfo {author} {\bibfnamefont {P.}~\bibnamefont {Coleman}}, \ and\
  \bibinfo {author} {\bibfnamefont {A.~M.}\ \bibnamefont {Tsvelik}},\ }\href
  {\doibase 10.1103/PhysRevLett.119.057603} {\bibfield  {journal} {\bibinfo
  {journal} {Phys. Rev. Lett.}\ }\textbf {\bibinfo {volume} {119}},\ \bibinfo
  {pages} {057603} (\bibinfo {year} {2017})}\BibitemShut {NoStop}%
\bibitem [{\citenamefont {Nozi\'eres}\ and\ \citenamefont
  {Blandin}(1980)}]{Nozieres_Blandin}%
  \BibitemOpen
  \bibfield  {author} {\bibinfo {author} {\bibfnamefont {P.}~\bibnamefont
  {Nozi\'eres}}\ and\ \bibinfo {author} {\bibfnamefont {A.}~\bibnamefont
  {Blandin}},\ }\href@noop {} {\bibfield  {journal} {\bibinfo  {journal} {J.
  Phys. France}\ }\textbf {\bibinfo {volume} {41}},\ \bibinfo {pages} {193}
  (\bibinfo {year} {1980})}\BibitemShut {NoStop}%
\bibitem [{\citenamefont {Ludwig}\ and\ \citenamefont {Affleck}(1991)}]{AL2}%
  \BibitemOpen
  \bibfield  {author} {\bibinfo {author} {\bibfnamefont {A.~W.~W.}\
  \bibnamefont {Ludwig}}\ and\ \bibinfo {author} {\bibfnamefont
  {I.}~\bibnamefont {Affleck}},\ }\href {\doibase 10.1103/PhysRevLett.67.3160}
  {\bibfield  {journal} {\bibinfo  {journal} {Phys. Rev. Lett.}\ }\textbf
  {\bibinfo {volume} {67}},\ \bibinfo {pages} {3160} (\bibinfo {year}
  {1991})}\BibitemShut {NoStop}%
\bibitem [{\citenamefont {Emery}\ and\ \citenamefont
  {Kivelson}(1992)}]{Emery-Kivelson92}%
  \BibitemOpen
  \bibfield  {author} {\bibinfo {author} {\bibfnamefont {V.~J.}\ \bibnamefont
  {Emery}}\ and\ \bibinfo {author} {\bibfnamefont {S.}~\bibnamefont
  {Kivelson}},\ }\href {\doibase 10.1103/PhysRevB.46.10812} {\bibfield
  {journal} {\bibinfo  {journal} {Phys. Rev. B}\ }\textbf {\bibinfo {volume}
  {46}},\ \bibinfo {pages} {10812} (\bibinfo {year} {1992})}\BibitemShut
  {NoStop}%
\bibitem [{\citenamefont {Jones}\ \emph {et~al.}(1988)\citenamefont {Jones},
  \citenamefont {Varma},\ and\ \citenamefont {Wilkins}}]{Jones_V_Wilkins}%
  \BibitemOpen
  \bibfield  {author} {\bibinfo {author} {\bibfnamefont {B.~A.}\ \bibnamefont
  {Jones}}, \bibinfo {author} {\bibfnamefont {C.~M.}\ \bibnamefont {Varma}}, \
  and\ \bibinfo {author} {\bibfnamefont {J.~W.}\ \bibnamefont {Wilkins}},\
  }\href {\doibase 10.1103/PhysRevLett.61.125} {\bibfield  {journal} {\bibinfo
  {journal} {Phys. Rev. Lett.}\ }\textbf {\bibinfo {volume} {61}},\ \bibinfo
  {pages} {125} (\bibinfo {year} {1988})}\BibitemShut {NoStop}%
\bibitem [{\citenamefont {Affleck}\ and\ \citenamefont
  {Ludwig}(1992)}]{AL3_two_imp1992}%
  \BibitemOpen
  \bibfield  {author} {\bibinfo {author} {\bibfnamefont {I.}~\bibnamefont
  {Affleck}}\ and\ \bibinfo {author} {\bibfnamefont {A.~W.~W.}\ \bibnamefont
  {Ludwig}},\ }\href {\doibase 10.1103/PhysRevLett.68.1046} {\bibfield
  {journal} {\bibinfo  {journal} {Phys. Rev. Lett.}\ }\textbf {\bibinfo
  {volume} {68}},\ \bibinfo {pages} {1046} (\bibinfo {year}
  {1992})}\BibitemShut {NoStop}%
\bibitem [{\citenamefont {Sire}\ \emph {et~al.}(1993)\citenamefont {Sire},
  \citenamefont {Varma},\ and\ \citenamefont {Krishnamurthy}}]{SVK}%
  \BibitemOpen
  \bibfield  {author} {\bibinfo {author} {\bibfnamefont {C.}~\bibnamefont
  {Sire}}, \bibinfo {author} {\bibfnamefont {C.~M.}\ \bibnamefont {Varma}}, \
  and\ \bibinfo {author} {\bibfnamefont {H.~R.}\ \bibnamefont
  {Krishnamurthy}},\ }\href {\doibase 10.1103/PhysRevB.48.13833} {\bibfield
  {journal} {\bibinfo  {journal} {Phys. Rev. B}\ }\textbf {\bibinfo {volume}
  {48}},\ \bibinfo {pages} {13833} (\bibinfo {year} {1993})}\BibitemShut
  {NoStop}%
\bibitem [{\citenamefont {Haldane}(1977)}]{Haldane1977_mixedval}%
  \BibitemOpen
  \bibfield  {author} {\bibinfo {author} {\bibfnamefont {F.~D.~M.}\
  \bibnamefont {Haldane}},\ }\href {\doibase 10.1103/PhysRevB.15.2477}
  {\bibfield  {journal} {\bibinfo  {journal} {Phys. Rev. B}\ }\textbf {\bibinfo
  {volume} {15}},\ \bibinfo {pages} {2477} (\bibinfo {year}
  {1977})}\BibitemShut {NoStop}%
\bibitem [{\citenamefont {Varma}(1994)}]{CMVcorrIns}%
  \BibitemOpen
  \bibfield  {author} {\bibinfo {author} {\bibfnamefont {C.~M.}\ \bibnamefont
  {Varma}},\ }\href {\doibase 10.1103/PhysRevB.50.9952} {\bibfield  {journal}
  {\bibinfo  {journal} {Phys. Rev. B}\ }\textbf {\bibinfo {volume} {50}},\
  \bibinfo {pages} {9952} (\bibinfo {year} {1994})}\BibitemShut {NoStop}%
\bibitem [{\citenamefont {Perakis}\ and\ \citenamefont
  {Varma}(1994)}]{PerakisV}%
  \BibitemOpen
  \bibfield  {author} {\bibinfo {author} {\bibfnamefont {I.~E.}\ \bibnamefont
  {Perakis}}\ and\ \bibinfo {author} {\bibfnamefont {C.~M.}\ \bibnamefont
  {Varma}},\ }\href {\doibase 10.1103/PhysRevB.49.9041} {\bibfield  {journal}
  {\bibinfo  {journal} {Phys. Rev. B}\ }\textbf {\bibinfo {volume} {49}},\
  \bibinfo {pages} {9041} (\bibinfo {year} {1994})}\BibitemShut {NoStop}%
\bibitem [{\citenamefont {Ruckenstein}\ and\ \citenamefont
  {Varma}(1991)}]{Ruck_V}%
  \BibitemOpen
  \bibfield  {author} {\bibinfo {author} {\bibfnamefont {A.}~\bibnamefont
  {Ruckenstein}}\ and\ \bibinfo {author} {\bibfnamefont {C.~M.}\ \bibnamefont
  {Varma}},\ }\href {https://doi.org/10.1016/0921-4534(91)91962-4} {\bibfield
  {journal} {\bibinfo  {journal} {Physica C}\ }\textbf {\bibinfo {volume}
  {185-189}},\ \bibinfo {pages} {134} (\bibinfo {year} {1991})}\BibitemShut
  {NoStop}%
\bibitem [{\citenamefont {Dzero}\ \emph {et~al.}(2010)\citenamefont {Dzero},
  \citenamefont {Sun}, \citenamefont {Galitski},\ and\ \citenamefont
  {Coleman}}]{Dzero_Coleman}%
  \BibitemOpen
  \bibfield  {author} {\bibinfo {author} {\bibfnamefont {M.}~\bibnamefont
  {Dzero}}, \bibinfo {author} {\bibfnamefont {K.}~\bibnamefont {Sun}}, \bibinfo
  {author} {\bibfnamefont {V.}~\bibnamefont {Galitski}}, \ and\ \bibinfo
  {author} {\bibfnamefont {P.}~\bibnamefont {Coleman}},\ }\href {\doibase
  10.1103/PhysRevLett.104.106408} {\bibfield  {journal} {\bibinfo  {journal}
  {Phys. Rev. Lett.}\ }\textbf {\bibinfo {volume} {104}},\ \bibinfo {pages}
  {106408} (\bibinfo {year} {2010})}\BibitemShut {NoStop}%
\bibitem [{\citenamefont {Park}\ \emph {et~al.}(2016)\citenamefont {Park},
  \citenamefont {Sun}, \citenamefont {Noddings}, \citenamefont {Kim},
  \citenamefont {Fisk},\ and\ \citenamefont {Greene}}]{Greene2016TopoSmB6}%
  \BibitemOpen
  \bibfield  {author} {\bibinfo {author} {\bibfnamefont {W.~K.}\ \bibnamefont
  {Park}}, \bibinfo {author} {\bibfnamefont {L.}~\bibnamefont {Sun}}, \bibinfo
  {author} {\bibfnamefont {A.}~\bibnamefont {Noddings}}, \bibinfo {author}
  {\bibfnamefont {D.-J.}\ \bibnamefont {Kim}}, \bibinfo {author} {\bibfnamefont
  {Z.}~\bibnamefont {Fisk}}, \ and\ \bibinfo {author} {\bibfnamefont {L.~H.}\
  \bibnamefont {Greene}},\ }\href {\doibase 10.1073/pnas.1606042113} {\bibfield
   {journal} {\bibinfo  {journal} {Proceedings of the National Academy of
  Sciences}\ }\textbf {\bibinfo {volume} {113}},\ \bibinfo {pages} {6599}
  (\bibinfo {year} {2016})},\ \Eprint
  {http://arxiv.org/abs/https://www.pnas.org/content/113/24/6599.full.pdf}
  {https://www.pnas.org/content/113/24/6599.full.pdf} \BibitemShut {NoStop}%
\bibitem [{\citenamefont {Nickerson}\ \emph {et~al.}(1971)\citenamefont
  {Nickerson}, \citenamefont {White}, \citenamefont {Lee}, \citenamefont
  {Bachmann}, \citenamefont {Geballe},\ and\ \citenamefont
  {Hull}}]{Geballe1971}%
  \BibitemOpen
  \bibfield  {author} {\bibinfo {author} {\bibfnamefont {J.~C.}\ \bibnamefont
  {Nickerson}}, \bibinfo {author} {\bibfnamefont {R.~M.}\ \bibnamefont
  {White}}, \bibinfo {author} {\bibfnamefont {K.~N.}\ \bibnamefont {Lee}},
  \bibinfo {author} {\bibfnamefont {R.}~\bibnamefont {Bachmann}}, \bibinfo
  {author} {\bibfnamefont {T.~H.}\ \bibnamefont {Geballe}}, \ and\ \bibinfo
  {author} {\bibfnamefont {G.~W.}\ \bibnamefont {Hull}},\ }\href {\doibase
  10.1103/PhysRevB.3.2030} {\bibfield  {journal} {\bibinfo  {journal} {Phys.
  Rev. B}\ }\textbf {\bibinfo {volume} {3}},\ \bibinfo {pages} {2030} (\bibinfo
  {year} {1971})}\BibitemShut {NoStop}%
\bibitem [{\citenamefont {Li}\ \emph {et~al.}(2014)\citenamefont {Li},
  \citenamefont {Xiang}, \citenamefont {Yu}, \citenamefont {Asaba},
  \citenamefont {Lawson}, \citenamefont {Cai}, \citenamefont {Tinsman},
  \citenamefont {Berkley}, \citenamefont {Wolgast}, \citenamefont {Eo},
  \citenamefont {Kim}, \citenamefont {Kurdak}, \citenamefont {Allen},
  \citenamefont {Sun}, \citenamefont {Chen}, \citenamefont {Wang},
  \citenamefont {Fisk},\ and\ \citenamefont {Li}}]{Li2014SmB6}%
  \BibitemOpen
  \bibfield  {author} {\bibinfo {author} {\bibfnamefont {G.}~\bibnamefont
  {Li}}, \bibinfo {author} {\bibfnamefont {Z.}~\bibnamefont {Xiang}}, \bibinfo
  {author} {\bibfnamefont {F.}~\bibnamefont {Yu}}, \bibinfo {author}
  {\bibfnamefont {T.}~\bibnamefont {Asaba}}, \bibinfo {author} {\bibfnamefont
  {B.}~\bibnamefont {Lawson}}, \bibinfo {author} {\bibfnamefont
  {P.}~\bibnamefont {Cai}}, \bibinfo {author} {\bibfnamefont {C.}~\bibnamefont
  {Tinsman}}, \bibinfo {author} {\bibfnamefont {A.}~\bibnamefont {Berkley}},
  \bibinfo {author} {\bibfnamefont {S.}~\bibnamefont {Wolgast}}, \bibinfo
  {author} {\bibfnamefont {Y.~S.}\ \bibnamefont {Eo}}, \bibinfo {author}
  {\bibfnamefont {D.-J.}\ \bibnamefont {Kim}}, \bibinfo {author} {\bibfnamefont
  {C.}~\bibnamefont {Kurdak}}, \bibinfo {author} {\bibfnamefont {J.~W.}\
  \bibnamefont {Allen}}, \bibinfo {author} {\bibfnamefont {K.}~\bibnamefont
  {Sun}}, \bibinfo {author} {\bibfnamefont {X.~H.}\ \bibnamefont {Chen}},
  \bibinfo {author} {\bibfnamefont {Y.~Y.}\ \bibnamefont {Wang}}, \bibinfo
  {author} {\bibfnamefont {Z.}~\bibnamefont {Fisk}}, \ and\ \bibinfo {author}
  {\bibfnamefont {L.}~\bibnamefont {Li}},\ }\href {\doibase
  10.1126/science:1250366} {\bibfield  {journal} {\bibinfo  {journal}
  {Science.}\ }\textbf {\bibinfo {volume} {346}},\ \bibinfo {pages} {1208 }
  (\bibinfo {year} {2014})}\BibitemShut {NoStop}%
\bibitem [{\citenamefont {Biswas}\ \emph {et~al.}(2014)\citenamefont {Biswas},
  \citenamefont {Salman}, \citenamefont {Neupert}, \citenamefont {Morenzoni},
  \citenamefont {Pomjakushina}, \citenamefont {von Rohr}, \citenamefont
  {Conder}, \citenamefont {Balakrishnan}, \citenamefont {Hatnean},
  \citenamefont {Lees}, \citenamefont {Paul}, \citenamefont {Schilling},
  \citenamefont {Baines}, \citenamefont {Luetkens}, \citenamefont {Khasanov},\
  and\ \citenamefont {Amato}}]{BiswasMusR}%
  \BibitemOpen
  \bibfield  {author} {\bibinfo {author} {\bibfnamefont {P.~K.}\ \bibnamefont
  {Biswas}}, \bibinfo {author} {\bibfnamefont {Z.}~\bibnamefont {Salman}},
  \bibinfo {author} {\bibfnamefont {T.}~\bibnamefont {Neupert}}, \bibinfo
  {author} {\bibfnamefont {E.}~\bibnamefont {Morenzoni}}, \bibinfo {author}
  {\bibfnamefont {E.}~\bibnamefont {Pomjakushina}}, \bibinfo {author}
  {\bibfnamefont {F.}~\bibnamefont {von Rohr}}, \bibinfo {author}
  {\bibfnamefont {K.}~\bibnamefont {Conder}}, \bibinfo {author} {\bibfnamefont
  {G.}~\bibnamefont {Balakrishnan}}, \bibinfo {author} {\bibfnamefont {M.~C.}\
  \bibnamefont {Hatnean}}, \bibinfo {author} {\bibfnamefont {M.~R.}\
  \bibnamefont {Lees}}, \bibinfo {author} {\bibfnamefont {D.~M.}\ \bibnamefont
  {Paul}}, \bibinfo {author} {\bibfnamefont {A.}~\bibnamefont {Schilling}},
  \bibinfo {author} {\bibfnamefont {C.}~\bibnamefont {Baines}}, \bibinfo
  {author} {\bibfnamefont {H.}~\bibnamefont {Luetkens}}, \bibinfo {author}
  {\bibfnamefont {R.}~\bibnamefont {Khasanov}}, \ and\ \bibinfo {author}
  {\bibfnamefont {A.}~\bibnamefont {Amato}},\ }\href {\doibase
  10.1103/PhysRevB.89.161107} {\bibfield  {journal} {\bibinfo  {journal} {Phys.
  Rev. B}\ }\textbf {\bibinfo {volume} {89}},\ \bibinfo {pages} {161107}
  (\bibinfo {year} {2014})}\BibitemShut {NoStop}%
\bibitem [{\citenamefont {Phelan}\ \emph {et~al.}(2014)\citenamefont {Phelan},
  \citenamefont {Koohpayeh}, \citenamefont {Cottingham}, \citenamefont
  {Freeland}, \citenamefont {Leiner}, \citenamefont {Broholm},\ and\
  \citenamefont {McQueen}}]{PhelanSphtSmB6}%
  \BibitemOpen
  \bibfield  {author} {\bibinfo {author} {\bibfnamefont {W.~A.}\ \bibnamefont
  {Phelan}}, \bibinfo {author} {\bibfnamefont {S.~M.}\ \bibnamefont
  {Koohpayeh}}, \bibinfo {author} {\bibfnamefont {P.}~\bibnamefont
  {Cottingham}}, \bibinfo {author} {\bibfnamefont {J.~W.}\ \bibnamefont
  {Freeland}}, \bibinfo {author} {\bibfnamefont {J.~C.}\ \bibnamefont
  {Leiner}}, \bibinfo {author} {\bibfnamefont {C.~L.}\ \bibnamefont {Broholm}},
  \ and\ \bibinfo {author} {\bibfnamefont {T.~M.}\ \bibnamefont {McQueen}},\
  }\href {\doibase 10.1103/PhysRevX.4.031012} {\bibfield  {journal} {\bibinfo
  {journal} {Phys. Rev. X}\ }\textbf {\bibinfo {volume} {4}},\ \bibinfo {pages}
  {031012} (\bibinfo {year} {2014})}\BibitemShut {NoStop}%
\bibitem [{\citenamefont {Wilson}(1975)}]{KWilson}%
  \BibitemOpen
  \bibfield  {author} {\bibinfo {author} {\bibfnamefont {K.~G.}\ \bibnamefont
  {Wilson}},\ }\href {\doibase 10.1103/RevModPhys.47.773} {\bibfield  {journal}
  {\bibinfo  {journal} {Rev. Mod. Phys.}\ }\textbf {\bibinfo {volume} {47}},\
  \bibinfo {pages} {773} (\bibinfo {year} {1975})}\BibitemShut {NoStop}%
\bibitem [{\citenamefont {Giamarchi}(2003)}]{Giamarchi}%
  \BibitemOpen
  \bibfield  {author} {\bibinfo {author} {\bibfnamefont {T.}~\bibnamefont
  {Giamarchi}},\ }\href@noop {} {\emph {\bibinfo {title} {Quantum Physics in
  One Dimension}}}\ (\bibinfo  {publisher} {Oxford University Press, Oxford},\
  \bibinfo {year} {2003})\BibitemShut {NoStop}%
\bibitem [{\citenamefont {Affleck}\ and\ \citenamefont {Ludwig}(1991)}]{AL1}%
  \BibitemOpen
  \bibfield  {author} {\bibinfo {author} {\bibfnamefont {I.}~\bibnamefont
  {Affleck}}\ and\ \bibinfo {author} {\bibfnamefont {A.~W.~W.}\ \bibnamefont
  {Ludwig}},\ }\href {\doibase 10.1103/PhysRevLett.67.161} {\bibfield
  {journal} {\bibinfo  {journal} {Phys. Rev. Lett.}\ }\textbf {\bibinfo
  {volume} {67}},\ \bibinfo {pages} {161} (\bibinfo {year} {1991})}\BibitemShut
  {NoStop}%
\bibitem [{\citenamefont {Taillefer}\ and\ \citenamefont
  {Lonzarich}(1988)}]{Lonzarich1988FS_UPt3}%
  \BibitemOpen
  \bibfield  {author} {\bibinfo {author} {\bibfnamefont {L.}~\bibnamefont
  {Taillefer}}\ and\ \bibinfo {author} {\bibfnamefont {G.~G.}\ \bibnamefont
  {Lonzarich}},\ }\href {\doibase 10.1103/PhysRevLett.60.1570} {\bibfield
  {journal} {\bibinfo  {journal} {Phys. Rev. Lett.}\ }\textbf {\bibinfo
  {volume} {60}},\ \bibinfo {pages} {1570} (\bibinfo {year}
  {1988})}\BibitemShut {NoStop}%
\bibitem [{\citenamefont {Toulouse}(1970)}]{Toulouse1970}%
  \BibitemOpen
  \bibfield  {author} {\bibinfo {author} {\bibfnamefont {G.}~\bibnamefont
  {Toulouse}},\ }\href {\doibase 10.1103/PhysRevB.2.270} {\bibfield  {journal}
  {\bibinfo  {journal} {Phys. Rev. B}\ }\textbf {\bibinfo {volume} {2}},\
  \bibinfo {pages} {270} (\bibinfo {year} {1970})}\BibitemShut {NoStop}%
\bibitem [{\citenamefont {Knolle}\ and\ \citenamefont
  {Cooper}(2015)}]{Knolle_Cooper}%
  \BibitemOpen
  \bibfield  {author} {\bibinfo {author} {\bibfnamefont {J.}~\bibnamefont
  {Knolle}}\ and\ \bibinfo {author} {\bibfnamefont {N.~R.}\ \bibnamefont
  {Cooper}},\ }\href {\doibase 10.1103/PhysRevLett.115.146401} {\bibfield
  {journal} {\bibinfo  {journal} {Phys. Rev. Lett.}\ }\textbf {\bibinfo
  {volume} {115}},\ \bibinfo {pages} {146401} (\bibinfo {year}
  {2015})}\BibitemShut {NoStop}%
\bibitem [{\citenamefont {Sodemann}\ \emph {et~al.}(2018)\citenamefont
  {Sodemann}, \citenamefont {Chowdhury},\ and\ \citenamefont
  {Senthil}}]{Chowdhury_Senthil}%
  \BibitemOpen
  \bibfield  {author} {\bibinfo {author} {\bibfnamefont {I.}~\bibnamefont
  {Sodemann}}, \bibinfo {author} {\bibfnamefont {D.}~\bibnamefont {Chowdhury}},
  \ and\ \bibinfo {author} {\bibfnamefont {T.}~\bibnamefont {Senthil}},\ }\href
  {\doibase 10.1103/PhysRevB.97.045152} {\bibfield  {journal} {\bibinfo
  {journal} {Phys. Rev. B}\ }\textbf {\bibinfo {volume} {97}},\ \bibinfo
  {pages} {045152} (\bibinfo {year} {2018})}\BibitemShut {NoStop}%
\bibitem [{\citenamefont {Varma}(1977)}]{cmvTmSe}%
  \BibitemOpen
  \bibfield  {author} {\bibinfo {author} {\bibfnamefont {C.}~\bibnamefont
  {Varma}},\ }\href@noop {} {\bibfield  {journal} {\bibinfo  {journal} {Solid
  State Comm.}\ }\textbf {\bibinfo {volume} {30}},\ \bibinfo {pages} {537}
  (\bibinfo {year} {1977})}\BibitemShut {NoStop}%
\end{thebibliography}

%

\end{document}